\documentclass[aps,prb,preprint,showpacs,floatfix]{revtex4}
\usepackage{amssymb}
\usepackage{graphicx}

\begin{document}

\title{Can nano-particle stand above
               the melting temperature of its fixed surface partner?}

\author{Yuanyuan Xu}
\affiliation{State Key Laboratory of Theoretical Physics,
         Institute of Theoretical Physics, Chinese Academy of Sciences,
         Beijing 100190, China }
\author{Kai Kang}
\affiliation{Science and Technology on Surface Physics
                  and Chemistry Laboratory, P.O. Box 718-35,
                             Mianyang 621907, Sichuan, China}
\author{Shaojing Qin}
\affiliation{State Key Laboratory of Theoretical Physics,
         Institute of Theoretical Physics, Chinese Academy of Sciences,
         Beijing 100190, China }

\date{\today}

\begin{abstract}


The phonon thermal contribution to the melting temperature of
nano-particles is inspected.
Unlike in periodic boundary condition,
under a general boundary condition the integration volume of
low energy phonon for a nano-particle is more complex.
We estimate the size-dependent melting temperature through
the phase shift of the low energy phonon mode
acquired by its scattering on boundary surface.
A nano-particle can have either a rising or a decreasing melting
temperature due to the boundary condition effect,
and we found that an upper melting temperature bound exists
for a nano-particle in various environments.
Moreover, the melting temperature under a fixed boundary condition sets
this upper bound.

\end{abstract}

\pacs{65.80.-g, 63.22.Kn, 81.70.Pg}

\maketitle

\section{INTRODUCTION}
\label{sec:intro}

This work focuses on the quantum size effect with particular emphasis
on environment-depending melting temperature of nano-particles.
The continuing progress in the design of nano-particles led to enhanced
and novel functionality \cite{STjong:2004}.
Concerning the thermal stability of heat resistant \cite{luk:2007},
we ask a question:
can nano-particles stand a temperature higher than the one its fixed
surface partners melt at?
There are many different thermodynamic theories of small systems
 \cite{jiang:2009,nanda:2012,yangcc:2014},
each considering different important aspects of size-dependent
melting of nano-particles.
The discrete quantum energy level has not been carefully considered
in these theories.
The critical role of phonon in thermal related phenomena is well-known
\cite{A-1,baowen:2006}.
But only recently the finite spacing energy levels became
a greater awareness in size-dependent melting \cite{SXH:2014}.
There is a chance to understand the rising of size-dependent melting
temperature of nano-particles with more attention on the important
aspect of quantum finite size effect.
We estimate in this work the change of melting temperature
by coating a fixed size particle or changing its environment.
We give the upper limit of the change on melting temperature.

The melting temperature for small particles was modeled and studied
in more than 100 years ago \cite{PPawlow:1909},
and the pressing need for
a deep understanding continues today \cite{C-1,C-2,C-3,C-4,C-5}.
In this study, we use the Lindemann melting
criterion \cite{FLindemann:1910,N-29,KK:2009} for its simplicity
in estimating the melting temperature of a nano-particle with
different boundary conditions.
By Lindemann criterion,
a nano-particle melts at the temperature $T_m$
at which the ratio of $u$, the square root of the mean square of
atom thermal displacement, to $a$ the lattice constant reaches the
Lindemann critical value $L_c$:
\begin{eqnarray}
L_c&=&\frac{u(T_m)}{a} . \label{eq:ldm}
\end{eqnarray}

\section{periodic boundary condition}

We repeat some necessary derivation for $u(T)$ in
harmonic approximation.
A lattice specified by a set of the vectors $\mathbf{R_i}$ will be
studied, with one atom at each lattice point.
The displacement $\mathbf{u}_i$ of an atom from its equilibrium
position $\mathbf{R}_i$ can be calculated by
\begin{eqnarray}
H&=&
{\textstyle \frac{M}{2}} \sum_{i\alpha}\dot{u}_{i\alpha}^2
+{\textstyle \frac{1}{2}} \sum_{ij\alpha\beta}
\Phi_{\alpha\beta}(\mathbf{R_i},\mathbf{R_j})u_{i\alpha}u_{j\beta},
\label{eq:ham}
\end{eqnarray}
where $\alpha=x, y, z$ and $u_{i\alpha}$ is the $\alpha$th
component of the displacement.
$\mathbf{R_i}$ $=$
$(i_x-\frac{L+1}{2},i_y-\frac{L+1}{2},i_z -\frac{L+1}{2})a$,
$i_\alpha=1, \ldots, L$.
The center of the nano-particle is at the origin of the coordinates,
and the boundary of the nano-particle is set on
$R_\alpha =\pm \frac{La}{2}$ surfaces in coordinate space.
$N=L^3$ is the number of atoms of the nano-particle.
$M$ is the mass of the atom.
The potential energy $\Phi$ is expanded to the second order and
$\Phi_{\alpha\beta}(\mathbf{R_i},\mathbf{R_j})$ $=$
$(\partial^2 \Phi/\partial u_{i\alpha}\partial u_{j\beta} )_0$.

The equations of motion of the lattice is then
\begin{eqnarray}
M\ddot{u}_{i\alpha}=-\sum_{j\beta}
\Phi_{\alpha\beta}(\mathbf{R_i},\mathbf{R_j})u_{j\beta} .
\label{eq:motion}
\end{eqnarray}
The general solution can be written in vibration modes $ Q_{k\sigma} $:
\begin{eqnarray}
u_{i\alpha}(t)={\textstyle \sqrt{\frac{1}{M}}} \sum_{k\sigma}
  Q_{k\sigma} e_{ k \sigma \alpha } e^{-i\omega_{\sigma}(\mathbf{k}) t}
\prod_{\alpha'} f(k_{\alpha'},R_{i\alpha'}) .
\label{eq:u_Q}
\end{eqnarray}
$\omega_{\sigma}(\mathbf{k})$ and $ \mathbf{e}_{ k \sigma } $
are phonon frequency and phonon polarization
of wave-vector $ \mathbf{k}$, respectively.
The atomic mean-squared thermal displacement is \cite{calaway}
\begin{eqnarray}
\langle u_{i\alpha}^2\rangle=\sum_{k\sigma}\frac{\hbar\,
e_{ k \sigma \alpha }^2} {NM\omega_{\sigma}(\mathbf{k})}
\left[\frac{1}{e^\frac{\hbar\omega_{\sigma}(\mathbf{k})}{k_B T}-1}
            +\frac{1}{2}\right],
\label{eq:ui2}
\end{eqnarray}
where $\langle\,\rangle$ means grand canonical ensemble average.
The center-of-mass motion is
$\mathbf{u}^{cm}=\frac{1}{N}\sum_i\mathbf{u}_i$.
The square root of the mean square of atom thermal
displacement, $u(T)$, is given by :
\begin{eqnarray}
u(T)&=&\sqrt{\frac{
\sum_{i\alpha} \left[\langle u_{i\alpha}^2\rangle -
\langle (u^{cm}_{\alpha})^2\rangle \right]}{N}}.
\label{eq:uT}
\end{eqnarray}
Melting temperature $T_m$ is obtained by solving Eq.~(\ref{eq:ldm}).
We will use boundary conditions to account for various environments.

For periodic boundary condition the mode expansion function for
atom displacement in each $\alpha$-direction is
$ f(k_{\alpha},R_{i\alpha}) $ $=$
$ \sqrt{\frac{1}{L}}e^{i k_{\alpha} R_{i\alpha}}$,
$k_{\alpha}= 2n_\alpha\pi/ L a$, and $n_\alpha = -L/2+1,\cdots,L/2$.
The melting temperature $T_{mn}$ for size $L$ nano-particle under
a periodic boundary condition is then calculated \cite{SXH:2014}:
\begin{eqnarray}
L_c^2&=& \frac{u^2}{a^2} \nonumber \\
     &=& \frac{3\hbar a }{16 \pi^3 M} \int_{-\pi/a}^{\pi/a} d^3 k
     \frac{1}{\omega(\mathbf{k})
     \tanh[\frac{\hbar\omega(\mathbf{k})}{2k_B T_{mn}}]}
\label{eq:tmpbc}
     \\
     & & -\frac{3\hbar a }{16 \pi^3 M}
  \int_{-\frac{\pi}{La}}^{\frac{\pi}{La}} d^3 k
          \frac{1}{\omega(\mathbf{k})
          \tanh[\frac{\hbar\omega(\mathbf{k})}{2k_B T_{mn}}]} ,
\nonumber
\end{eqnarray}
where ${\omega(\mathbf{k})}={\omega_{\sigma}(\mathbf{k})}$ is used
when summing up the three polarized vibration directions
in each $\mathbf{k}$ mode.
$\mathbf{u}^{cm}$ is given only by the zero wave-vector phonon at $k=0$.
The second term removes contribution of the global moving $(u^{cm})^2$.
This is the missing of phonon contribution from the zero-mode volume.
We have repeated briefly the derivation of this zero-mode volume
in the work of Sui et al\cite{SXH:2014}.
For the bulk system, melting temperature $T_{mb}$ is the solution
of the same equation in the limit of $1/L \to 0$,
which turns the second term into zero:
\begin{equation}
L_c^2 = \frac{3\hbar a }{2 \pi^3 M}
     \int_{0}^{\pi/a} d^3 k \frac{1}{\omega(\mathbf{k})
     \tanh[\frac{\hbar\omega(\mathbf{k})}{2k_B T_{mb}}]}
      .
\label{eq:tmbulk}
\end{equation}

\section{boundary reflective phase shift}

We analyze the phase shift of the low energy phonon mode for
different boundary conditions.
A theoretical estimated expression for the melting temperature of
nano-particle will contain physical quantities with some uncertainty.
Beside the main variable, the size $L$, other physical factors may
play a role in finite size melting, such as shape, surface
reconstruction, and environmental effect etc.
An expression has to have physical quantities not so certain
to absorb all the different factors from material to material,
environment to environment, and particle to particle. We will use
the phase shift of low energy acoustic phonon for this function.
A model with boundary is good for nano-particles under different
boundary conditions.
The phase shift of phonon mode depends on its scattering on
the surface of a nano-particle.
A phase shift can be drawn from the
experimental data, or it has to be chosen with some uncertainty
for a particular situation before we can make estimation for
a design of the melting temperature.

The scattering and phase shift have been studied most thoroughly in the
context of quantum theory. The notion of reflection of waves in
one-dimensional scattering plays a central role in the following
detailed discussion. We model the boundary of the nano-particle
as an additional potential which is bigger than zero outside the
nano-particle and zero inside the particle.
In any $\alpha$-direction, the general solution
for Eq.~(\ref{eq:motion}) is in terms of waves
moving in opposite directions, $f(k_{\alpha},R_{i\alpha})$ $=$
$c ~  e^{i k_\alpha R_{i\alpha}} + d ~ e^{- i k_\alpha R_{i\alpha}}$,
for each vibration mode $ Q_{k\sigma} $.  We study the standing waves
under a general real boundary potential.  The most general form for
the reflection coefficient would then be $r=e^{i\delta_{k\alpha}}$;
the termination at the boundary could at most introduce a phase
change in the reflected wave:
\begin{eqnarray}
\frac{d ~  e^{-i k_\alpha R_{\alpha}}}{c ~  e^{i k_\alpha R_{\alpha}}}
   = e^{i\delta_{k\alpha}}, ~~~~~
 R_{\alpha} = La/2.
\label{eq:reflect}
\end{eqnarray}

Different boundary conditions resulting in the same magnitude of
phase shift are physically equivalent.
A very careful but long and technical analysis on phase shift and
boundary condition can provide considerable understanding of
nano-particle's surface and environment.
We will put this study under control.
We assume an isotropic surface and environment.
The same phase shift will be assumed in each $\alpha$-direction.
The phase shift for a small wave-vector is expanded up to
the first order of $1/L$: $\delta_{k\alpha}=\delta+a_1 k_\alpha$.
$\delta$ is a constant phase shift and
$a_1$ is the expansion coefficient for the first order term.
We will discuss on phase shift in the range of $[-\pi,0]$,
with corresponding boundary barrier effectively repulsive.
This phase shift is the parameter we used to
model the boundary effect.

The wave-vector is fixed by the boundary
condition Eq.~(\ref{eq:reflect}).
First, when the standing wave is of even parity, we have $c/d=1$ in
$f(k_{\alpha},R_{i\alpha})$,
and  $e^{i(k_{\alpha} La + \delta_{k\alpha})} =1$.
The mode expansion for atom displacement on this kind of phonon mode is
$f(k^e_{\alpha},R_{i\alpha})$ $\sim$ $\cos(k^e_{\alpha} R_{i\alpha})$,
with $k^e_\alpha = \frac{2 n_\alpha \pi + |\delta_{k^e\alpha}|}{La}$.
The integer $n_\alpha$ runs from $0$ to $L/2-1$.
The second group is the odd parity group for $c/d=-1$ in
$f(k_{\alpha},R_{i\alpha})$,
and  $e^{i(k_{\alpha} La + \pi + \delta_{k\alpha})} =1$.
The mode expansion for atom displacement on this kind of phonon mode is
$f(k^o_{\alpha},R_{i\alpha})$ $\sim$ $\sin(k^o_{\alpha} R_{i\alpha})$,
with $k^o_\alpha = \frac{(2 n_\alpha-1) \pi + |\delta_{k^o\alpha}|}{La}$.
The integer $n_\alpha$ runs from $1$ to $L/2$.
The low energy wave-vector has an increase $|\delta_{k\alpha}|/La$ when
$\delta_{k\alpha}\ne 0$.

\section{Fixed Boundary Condition}

With the above inspection, we will be able to estimate the melting
temperature of a nano-particle under a general boundary condition.
The low energy wave-vector has an increase $|\delta_{k\alpha}|/La$.
The melting temperature is increased by a boundary
condition shifting $k_\alpha$ wave-vectors upward.
The discrete summation of phonon modes in Eq.~(\ref{eq:ui2})
can be written into
$\int_{(2|\delta|-\pi)/2La} d k_\alpha$
for each $\alpha$-component of wave-vector \cite{SXH:2014}.
When $|\delta_{k\alpha}|=\pi/2$ the integration volume for
low energy phonon starts at $k_\alpha=0$:
$\int_{0} d k_\alpha$, which results in the same amount of
low energy phonon contribution to atom displacement as
the case for bulk material in Eq.~(\ref{eq:tmbulk}).
The cases for $|\delta_{k\alpha}| < \pi/2$ will depress
the melting temperature of a nano-particle.
The lower bound of this depression
when $|\delta_{k\alpha}| \to 0$ was carefully studied in
the work of Sui et. al \cite{SXH:2014}.
When $|\delta_{k\alpha}| > \pi/2$, the melting temperature of
a nano-particle will be higher than its bulk parent.

We study the rising of the melting temperature due to the phase shift.
The boundary scattering shifts wave-vectors up to
$k_\alpha=(|\delta_{k\alpha}| + n_\alpha \pi)/La$
by a nonzero $\delta_{k\alpha}$.
The range of the phase shift $\delta_{k\alpha}$ in $k_\alpha$
is $[- \pi,0]$.
If $|\delta_{k\alpha}|=\pi/2$, the density of the low energy
phonon states in a nano-particle is the same as the density in
the bulk material.
We prefer a smaller thermal displacement of the atom when we study
a heat resistant. We want less phonon
contribution to atom displacement. The least contribution from
phonon modes occurs when $\delta_{k\alpha} = -\pi$.
This situation is given by the ideal fixed boundary condition
(fixed-BC), for which the boundary sets up a node surface for
standing acoustic waves.

The upper melting temperature bound is set by the ideal fixed-BC
for nano-particles.
The phonon wave-vectors move up for this boundary condition
and some more low energy phonon contribution is taken away by
the deduction of the second $(u^{cm})^2$ term.
In the picture of microscopic physics, less low energy phonon will
contribute to atom displacements in finite size fixed-BC particle
than in bulk material.
Therefore, the melting temperature increases for the boundary effect
and the finite size effect.
Moreover, for all possible boundary conditions modeled in phase shifts,
the low energy phonon modes shift up the most for fixed-BC,
and their contribution missed the most for fixed-BC.
Thus the melting temperature increases the most for fixed-BC.

The upper limit for effective phase shift is $|\delta_{k\alpha}|=\pi$,
and the melting temperature increases at most when a particle
subject to a fixed boundary condition.
However, this ideal boundary condition is not easy to achieve when
preparing nano-powder heat resistant, since there is no infinitely
heavy walls we can coat microscopically on a nano-particle.
Fixed-BC is an unrealistic boundary condition but it is instructive.
Nano-particles can be grown inside sieves,
can be coated with other elements.
The melting temperature of a nano-particle can approach
but not reach the one for fixed-BC. From the above discussion on
both finite size and boundary condition, we can conclude that
the upper melting temperature bound is set by the ideal fixed-BC
for nano-particles.

We derive this upper melting temperature bound in the following.
For fixed-BC, the mode expansion for atom displacement
is $f(k^e_{\alpha},R_{i\alpha})$ $=$
$\sqrt{\frac{2}{L}}\cos(k^e_{\alpha} R_{i\alpha})$ for even parity,
and $f(k^o_{\alpha},R_{i\alpha})$ $=$
$\sqrt{\frac{2}{L}}\sin(k^o_{\alpha} R_{i\alpha})$ for odd parity.
Quite different from periodic condition's mode space,
there is no $k=0$ phonon mode for fixed-BC.
Even parity $k^e_\alpha$ will contribute to $\mathbf{u}^{cm}$,
since
$|\sum_{i\alpha=1}^{L} \cos[k^e_\alpha a (i_\alpha-\frac{L+1}{2}]|$ $=$
$1 / \sin\frac{k_\alpha^e a}{2}$.
The melting temperature $T_{mn}$ for size $L$ nano-particle under
a fixed-BC is then calculated from Eq.~(\ref{eq:ldm}):
\begin{eqnarray}
&L_c^2&= \frac{3\hbar }{2 a^2 N M} \left[
\sum_{k_x,k_y,k_z}
   \frac{1}{\omega(\mathbf{k})
   \tanh[\frac{\hbar\omega(\mathbf{k})}{2k_B T_{mn}}]} \right.
\label{eq:tmbc}
\\
   & - &
\left.
\sum_{k^e_x,k^e_y,k^e_z}
   \frac{8}{\omega(\mathbf{k})
   \tanh[\frac{\hbar\omega(\mathbf{k})}{2k_B T_{mn}}]
   \prod_\alpha { [L \sin\frac{k_\alpha^e a}{2}]^2 }
} \right] .
  \nonumber
\end{eqnarray}
The first term is the summation of $u_{i\alpha}^2$ defined by
Eq.~(\ref{eq:ui2}),
in which $k_\alpha=n_\alpha \pi/La$ with $n_\alpha=1,\cdots,L$.
The second term removes the contribution of the global
moving $(u^{cm})^2$,  which sums only the even parity
$k^e_\alpha=(2 n_\alpha+1) \pi/La$ with $n_\alpha=0,\cdots,L/2-1$.

\section{Discussion and application}

State-of-art functional designs of nano-particles developed
in recent years \cite{A-9,kalele:2006}.
The first general factor came into our understanding on size-dependent
melting was the surface volume ratio of number of atoms \cite{NN-nm}.
Nano-materials are studied case by case in experiments and Ab
initio calculations \cite{N-2,A-2,zhong:2001,dick:2002,rosner:2003}.
With the help of computer simulation, in principle all the
parameters can be fixed to certain extent for a particular
design of a particular particle, of its surface and its properties
\cite{A-4,storozhev:2010}.
A designed melting temperature could be accurate up to 1 K \cite{T-3}.
Mechanical stable region for nano-materials is also under current
studying along with the estimation of $T_{mn}$, and giving
the safe working temperature \cite{fischer:2008}.
Many on going theoretical researches are working for more general
understanding on size-dependent melting.

Statistical mechanics is valid for an ensemble of big
systems and small systems.
Thermodynamics of small systems is well understood.
New and powerful expressions were derived through
thermodynamics \cite{A-6} for size- and boundary-dependent melting,
but they have not converged to physically equivalent ones yet.
Size-effect on intensive variables can be included
through thermodynamic equations or ensemble average
of quantum states.
In this study on general boundary conditions for nano-particles,
it is clear that the boundary condition and size together determine
the wave-vectors of phonon modes, and so determine
the low energy density of states of phonon.
Such a size- and boundary-dependent density of states was not
included in previous approximations in a phonon Debye temperature.
In real applications, a particle interacts with other surrounding
atoms, molecules, and particles simultaneously.
Our analysis is right when the bulk phonon dispersion is almost kept
in a nano-particle \cite{cqsun:2006}.
The individuality
of the particle is retained under the statistical average.
Vibration modes inside the particle see only an averaged reflective
phase shift.

\begin{figure}
\includegraphics[width=8cm]{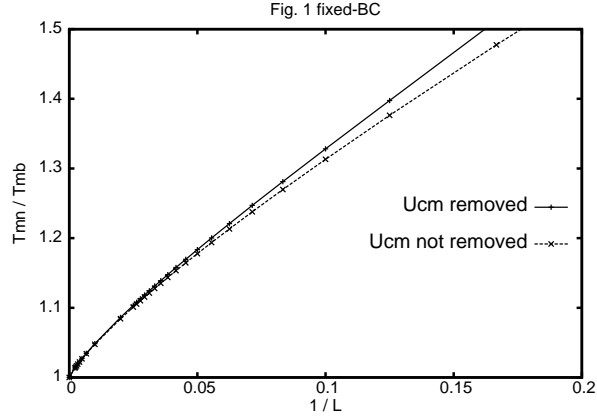}
\caption{
The size dependent melting temperature for nano-particles under fixed-BC.
The lines and points are $T_{mn}/T_{mb}$ with and without
the second term in Eq.~(\ref{eq:tmbc}).
Guiding lines simply connecting the points are presented.
}
\label{fig:fig1}
\end{figure}

In this study, a simple procedure to estimate the melting temperature
for nano-particles under a general boundary condition can be suggested.
In this simple estimation procedure we use sound speed $v$ to
approximate the phonon dispersion for all phonon modes
in Eq.~(\ref{eq:tmbulk}) for an integration, and
in Eq.~(\ref{eq:tmbc}) for a discrete summation.
For example,
in Fig. 1 we plot the size-dependent melting for a simple cubic
lattice under fixed-BC. Phonon dispersion is approximated by
$ \omega(\mathbf{k})$ $=$
$\frac{ 2 v }{a} \sqrt{ \sin^2 \frac{k_x a}{2}
+ \sin^2 \frac{k_y a}{2} + \sin^2 \frac{k_z a}{2} } $.
We substitute $\hbar v / a k_B T_{mb} $ by $0.25$ when solving
Eq.~(\ref{eq:tmbulk}) and Eq.~(\ref{eq:tmbc})
for $T_{mn}/T_{mb}$ at each size $L$.

Fig. 1 shows that the global movement contributes more
in the variation of melting temperature as the size decreases down
to $L=20$ and smaller.
The global movement deduction accounts for five percent of the change
when $L=10$.
For a nano-particle at the size of $L=20$,
with $\hbar v / a k_B T_{mb} \sim 0.25$ for its parent bulk material,
we can read on Fig. 1 that an increase of melting temperature
less than 18 percent can be expected
when we design a heat resistant coating for the nano-particle.
This example shows that the reflective phase shift at the particle
boundary plays an important role in size-dependent melting,
and size and phase shift are both general factors in this phenomenon.

\section{CONCLUSIONS}

We investigated in this paper the melting temperatures for
nano-particles under a general boundary condition.
We find that the missed contribution from the low energy phonon
will raise the melting temperature. This phenomenon comes from finite
size effect, and boundary condition plays an important role in it.
The melting temperature for the fixed boundary condition gives
the upper bound for the melting temperature of a nano-particle
in all kinds of environment. This upper bound is not reachable
in any real nano-particle design.

\section*{Acknowledgements}
The authors thank Xiaohong Sui and visitors to KITPC for discussions.
This work was supported by NNSF1121403 of China.

\end{document}